\documentclass[aps,prb,twocolumn,a4paper,floatfix,showpacs]{revtex4}
\usepackage{graphicx} 
\usepackage{times}
\usepackage{bm}
\usepackage{amssymb,amsmath}
\usepackage{ifthen}
\newboolean{JCPSub}
\setboolean{JCPSub}{false}

\newcommand{\mr}[1]{\mathrm{#1}}

\newcommand{\eps}{\varepsilon}
\newcommand{\half}{\frac{1}{2}}
\newcommand{\Def}{:=}
\newcommand{\br}{\mathbf{r}}

\newcommand{\Exc}{E_\mr{xc}}
\newcommand{\Ec}{E_\mr{c}}

\newcommand{\EcRPA}{\Ec^{\mr{RPA}}}

\newcommand{\vxc}{v_\mr{xc}}
\newcommand{\vs}{v_\mr{s}}

\newcommand{\vc}{v_\mr{c}}

\newcommand{\vint}{v_\mr{ee}}

\newcommand{\fxc}{f_\mr{xc}}

\newcommand{\TEI}[2]{\left( #1 \mid\mid #2\right)}

\newcommand{\Tr}[1]{ \mathrm{Tr} \left\lbrace #1 \right\rbrace }
\newcommand{\Ln}[1]{ \mathrm{Ln} \left(  #1 \right) }
\newcommand{\Det}[1]{ \mathrm{Det} \left(  #1 \right) }

\newcommand{\Imax}{I_\mathrm{max}}
\newcommand{\Rmax}{R_\mathrm{max}}
\newcommand{\nmax}{n_\mathrm{max}}
\newcommand{\lmax}{l_\mathrm{max}}
\newcommand{\emax}{\varepsilon_\mathrm{max}}

\newcommand{\ThreeJ}[6]
{  \left(
   \begin{array}{ccc} 
      #1 & #2 &  #3 \\ 
      #4 & #5 &  #6 
   \end{array} 
   \right)
}

\begin{document}
\title{%
Random-phase-approximation-based correlation energy functionals:
Benchmark results for atoms} 

\author{Hong Jiang }
\thanks{Current address: Fritz-Haber-Institut der Max-Planck-Gesellschaft, Faradayweg 4-6, D-14195 Berlin, Germany}
\author{Eberhard Engel}
\affiliation{Center for Scientific Computing, 
J.W.Goethe-Universit\"at Frankfurt, 
Max-von-Laue-Stra\ss{}e 1, 
D-60438 Frankfurt/Main, Germany}

\pacs{31.15.Ew, 31.10.+z, 71.15.-m}

\begin{abstract}
The random phase approximation (RPA) for the correlation energy
functional of density functional theory has recently attracted 
renewed interest.
Formulated in terms of the Kohn-Sham (KS) orbitals and eigenvalues, it 
promises to resolve some of the fundamental limitations of the local 
density and generalized gradient approximations, as for instance their
inability to account for dispersion forces.
First results for atoms, however, indicate that the RPA overestimates
correlation effects as much as the orbital-dependent functional 
obtained by a second order perturbation expansion on the basis of 
the KS Hamiltonian.
In this contribution, three simple extensions of the RPA are examined,
(a) its augmentation by an LDA for short-range correlation,
(b) its combination with the second order exchange term, and
(c) its combination with a partial resummation of the perturbation
series including the second order exchange.
It is found that the ground state and correlation energies as well as 
the ionization potentials resulting from the extensions (a) and (c) 
for closed sub-shell atoms are clearly superior to those obtained with 
the unmodified RPA.
Quite some effort is made to ensure highly converged RPA data, so that
the results may serve as benchmark data.
The numerical techniques developed in this context, in particular
for the inherent frequency integration, should also be useful for 
applications of RPA-type functionals to more complex systems. 
\end{abstract}
\maketitle

\section{Introduction}

Recent years have seen a revival of interest in the random phase 
approximation (RPA) and its extensions, both in the framework of 
Kohn-Sham density functional theory (KS-DFT) 
\cite{Dobson94,Pitarke98,Dobson99,Kurth99,Lein99,Yan00,Lein00,%
Dobson00,Furche01,Fuchs02,Miyake02,Fuchs03,Pitarke03,Jung04,%
Furche05,Fuchs05,Marini06} and within Green's function-based
many-body theory for ground state properties.\cite{SanchezFriera00,%
GarciaGonzalez02,Aryasetiawan02,Dahlen04b,Dahlen06} 
Within KS-DFT, the RPA for the energy and response function of the 
homogeneous electron gas played an important role in the development 
of the local-density approximation (LDA) as well as the generalized 
gradient approximation (GGA) for the exchange-correlation (XC) energy 
functional $E_{xc}$.\cite{ParrYang89,DreizlerGross90} 
Current interest in the RPA is stimulated by the concept of 
orbital-dependent (implicit) XC functionals, in which the XC energy 
is represented in terms of the KS orbitals and eigenenergies.%
\cite{Engel99,Grabo99,Engel03,Goerling05,Baerends05}
Within this approach an RPA-type correlation energy functional is most 
easily formulated on the basis of the KS response function.
Compared to LDA/GGA-type explicit XC functionals, implicit functionals 
have several attractive features:
(1) the exchange can be treated exactly, leading to exchange energies 
and potentials which are free of self-interaction;\cite{Talman76}
(2) the long-range dispersion interaction can be correctly described;
\cite{Dobson94, Engel98, Engel00a, Marini06}
(3) static correlation effects can be incorporated even within a 
spin-unpolarized formalism.\cite{Fuchs03} 

A systematic formulation of orbital-dependent XC functionals is 
possible within KS-based many-body theory, i.e.\  by using the KS 
Hamiltonian as non-interacting reference Hamiltonian in the framework 
of standard many-body theory (KS-MBT).\cite{Sham85,Goerling94,Engel03} 
In this approach the exact exchange of DFT emerges as first order
contribution to a straightforward perturbation expansion in powers 
of $e^2$.
All higher order terms constitute the DFT correlation energy. 
The lowest order correlation contribution resulting from perturbation 
theory, $E_c^{(2)}$, has been extensively studied for atoms and small 
molecules.\cite{Engel98,Engel00a,FaccoBonetti01,Grabowski02,%
MoriSanchez05,Jiang05} 
$E_c^{(2)}$ correctly accounts for the dispersion interaction 
\cite{Engel98,Engel00a} and the corresponding correlation potential 
$v_c^{(2)}$ reproduces the shell-structure and asymptotic behavior
of atomic correlation potentials.\cite{Jiang05} 
On the other hand, the magnitude of the energies and potentials 
resulting from $E_c^{(2)}$ overestimates the corresponding exact
data significantly.
Moreover, $E_c^{(2)}$ is found to be variationally instable for systems 
with a very small energy gap between the highest occupied and the 
lowest unoccupied molecular orbital (HOMO-LUMO gap) 
\cite{Jiang05, MoriSanchez05} (as, for instance, the 
beryllium atom) and fails to describe chemical bonding in such 
elementary molecules as the nitrogen dimer.\cite{Engel03}  
The variational instability of $E_c^{(2)}$ can be resolved by 
resummation of suitable higher order contributions to infinite order
(e.g.\  in the form of Feynman diagrams).
The simplest functional of this type is obtained by resummation of 
selected ladder-type diagrams, i.e.\  the Epstein-Nesbet(EN)-diagrams.
The resulting functional is not only found to be variationally stable 
for all neutral and singly-ionized atoms, but also gives more accurate 
correlation energies and potentials than $E_c^{(2)}$.\cite{Jiang06} 
However, EN-type functionals still face fundamental problems in the 
case of degenerate or near-degenerate systems.
A more suitable partial resummation scheme is needed to establish a 
universally applicable, implicit XC functional, the RPA and its 
extensions being the prime candidates.

In standard many-body theory, the RPA is obtained by resummation of 
the so-called ring diagrams.\cite{FetterWalecka}
This concept can be directly transfered to the framework of KS-MBT.%
\cite{EF01}
On the other hand, in the context of DFT, the RPA can also be derived 
from the adiabatic connection fluctuation-dissipation (ACFD) theorem.%
\cite{Langreth77} 
The ACFD formalism is, for instance, the conceptual starting point for 
the recent development of van-der-Waals DFT.\cite{Dobson94} 
It has also been applied directly to various systems, including jellium 
surfaces and slabs,\cite{Dobson99}, atoms \cite{Dahlen04b,Hellgren07}, 
small molecules \cite{Furche01,Fuchs02,Furche05,Dahlen06} and solids.%
\cite{Miyake02,Marini06} 
All these calculations have demonstrated promising features of 
RPA-based functionals. 
On the other hand, the results for atoms,\cite{Dahlen04b,Hellgren07} 
for which rigorous benchmark data are available, indicate that the 
pure RPA overestimates correlation energies and potentials as much 
as $E_c^{(2)}$.

One is therefore led to consider extensions of the RPA.
The most obvious starting point for extension is the inclusion of
the second order exchange (SOX) contribution.
However, in its pure form it neglects the screeening of the Coulomb
interaction, which is the core feature of the RPA.
One would thus expect an imbalance between direct and exchange 
contributions, when combining the RPA with the pure SOX term.
A fully screened form of the SOX is easily formulated, following
the line of thought used for the derivation of GGAs.\cite{Hu86}
The resulting functional, however, is computationally much more
demanding than the RPA.
For that reason it is worthwhile to examine alternative modifications 
of the SOX term which reduce its net contribution.
Given the success of the EN-resummation in the context of the complete
$E_c^{(2)}$, an EN-extension of the SOX term suggests itself
(this functional is denoted as RSOX in the following).

The SOX term, be it screened or not, is inherently a short-range
contribution.
This raises the question whether it is sufficient to account for the 
complete screened SOX in an approximate fashion, relying on the LDA.
In fact, using this strategy, one can easily include all short-range
correlation effects beyond the RPA.\cite{Kurth99,Yan00}
Clearly, the resulting LDA-type functional (here labelled as RPA+) is 
even more efficient than the RSOX. 

In this work, we study the RPA and these simple extensions for a series 
of prototype atoms and ions, for which highly accurate reference data 
are available for comparison.
In order to provide benchmark results a numerically exact, i.e.\
basis-set-free, approach is used and considerable emphasis is placed 
on all convergence issues involved.
As a byproduct of this strive for accuracy, a highly efficient scheme 
for performing the frequency integration inherent in all RPA-type 
functionals has been developed.
This procedure should be useful for applications to more complex 
systems, for which utilizing more than the minimum number of grid 
points for the frequency integration would be too demanding.  

A complete implementation of any XC functional requires not only 
the evaluation of the XC energy, but also the inclusion of the 
corresponding XC potential $\vxc$ in the self-consistent calculation.
The latter step is quite challenging in the case of orbital-dependent 
XC functionals, for which $\vxc$ has to be determined indirectly via 
the Optimized Potential Method (OPM),%
\cite{Talman76,Engel99,Grabo99,Engel03,Goerling05} 
and, in particular, for RPA-type functionals.\cite{Godby87,Kotani98,%
FaccoBonetti01,FaccoBonetti03,Niquet03b,Niquet03c,Jiang05,Gruening06}
Recently, Hellgren and von Barth have reported the first self-consistent
RPA correlation potentials for spherical atoms, obtained by solution of
the linearized Sham-Schluter equation \cite{Sham83} at the GW level.%
\cite{Hellgren07}
However, as indicated earlier, the RPA is not consistently improving
atomic correlation potentials over $E_c^{(2)}$.
In the present work we therefore focus on the perturbative evaluation
of all RPA-type energies, utilizing self-consistent exchange-only 
orbitals and eigenvalues. 
As we will show, the RPA correlation energy is rather insensitive 
to the KS orbitals used for its evaluation, which clearly supports
this perturbative approach.
This feature, if true in general, will be very important for the 
application of RPA-type functionals to more complicated systems,
for which a self-consistent implementation is not feasible anyway.

The paper is organized as follows. 
In the Section II, first the RPA correlation energy is formulated in 
the framework of the ACFD formalism. 
In addition, the general result is reduced to an expression valid for 
spherical systems.
In Section III, various numerical aspects are discussed, addressing
in particular questions of accuracy.
In Section IV, the RPA correlation energies for a number of prototype 
atoms and ions (we will no longer distinguish between neutral atoms 
and atomic ions in the following) are presented and compared to the 
corresponding exact data. 
Section V provides a summary. 
Atomic units are used throughout this paper.



\section{Theory}
\subsection{RPA correlation energy on basis of the ACFD formalism}
Based on the adiabatic connection and the zero-temperature 
fluctuation-dissipation theorem, \cite{Langreth75, Langreth77} the 
exact KS correlation energy can be written as
\begin{eqnarray}
E_{\rm c} 
&=& 
- \frac{1}{2 \pi} \int_0^\infty du \int_0^1 d\lambda \int d\br d\br'  \vint(\br-\br') \nonumber \\
&&\times   
\left[  \chi_\lambda(\br,\br';iu)- \chi_0(\br,\br';iu) \right]  
\label{eq:EcACFD}
\;,
\end{eqnarray} 
where $\vint(\br,\br') = 1/|\br-\br'|$ is the bare Coulomb interaction,
$\chi_0$ is the KS response function,
\begin{equation}
\chi_0(\br,\br'; iu ) 
=  
\sum_{ia} \frac{\phi_i^\dagger(\br) \phi_a(\br)
                \phi_a^\dagger(\br') \phi_i(\br')} 
               {iu + \eps_i - \eps_a} + c.c.,
\label{eq:Chi0}
\end{equation}
and $\chi_\lambda$, with $\lambda \in [0,1]$, is the density-density
response function of a fictitious system in which electrons interact 
with a scaled Coulomb potential $ \lambda \vint(\br,\br') $, and
simultaneously move in a modified external potential, chosen such that 
the electron density remains identical to that of the fully interacting
system with $\lambda=1$. 
Throughout this paper we use the convention that $i,j,\dots$ denote 
occupied (hole) KS states, while $a,b,\dots$ are used for unoccupied 
(particle) states and $p,q,\dots$ for the general case. 
$\chi_\lambda$ is related to $\chi_0$ by a Dyson-like integral equation,
\cite{Gross85}
\begin{widetext}
\begin{equation}
\chi_\lambda(\br,\br',iu) 
= 
\chi_0(\br,\br',iu) + \int d\br_1 \int d\br_2 \chi_0(\br,\br_1,iu) K_\lambda(\br_1,\br_2,iu) \chi_\lambda(\br_2,\br',iu)  
\label{eq:Dyson}
\;,
\end{equation}
\end{widetext}
where
\begin{equation}
K_\lambda(\br_1,\br_2,iu)
=
\lambda \vint(\br_1,\br_2) + \fxc^\lambda(\br_1,\br_2,iu)
\end{equation}
is the Coulomb and XC kernel. 

The RPA correlation energy is obtained from Eq.(\ref{eq:EcACFD}) if 
one neglects the XC contribution to the right-hand side of 
Eq.(\ref{eq:Dyson}). 
Integrating over $\lambda$ one ends up with 
\begin{equation} 
\EcRPA 
= 
\int_0^\infty \frac{du}{2\pi} \Tr{ \Ln{ 1-\chi_0(iu) \vint } + \chi_0(iu) \vint } 
\label{eq:EcRPA}
\;,
\end{equation}
where the trace indicates integration over all spatial coordinates. 
It is often more instructive to rewrite the integrand in Eq. 
(\ref{eq:EcRPA}), denoted as $E_c(iu)$, as a power series in the
Coulomb interaction,
\begin{eqnarray}
  E_c(iu) = -\sum_{n=2}^\infty \Tr{(\chi_0(iu) \vint)^n }  
\label{eq:EcRPA-2}
\;.
\end{eqnarray}

\subsection{Correlation energy beyond RPA}
The ACFD theorem provides a natural starting point for the development
of correlation functionals beyond the RPA:
Inclusion of some approximation for $f_{xc}$ in the Dyson equation 
(\ref{eq:Dyson}) automatically yields an extension of the RPA.
Several approximate XC kernels have been introduced in the context of 
time-dependent DFT (TDDFT).\cite{Gross96,Onida02}
It is, however, not clear whether an approximate $f_{xc}$ designed to
provide a good description of excited states within TDDFT also leads to 
accurate ground state correlation energies. 

A more straightforward extension of the RPA, the so-called RPA+ 
approach, had been proposed by Perdew and coworkers.%
\cite{Kurth99, Yan00} 
They observed that the RPA provides a quite accurate description of 
long-range correlation, but is inadequate for short-range correlations.
On the other hand, the latter can be very well approximated by a local 
or semi-local density-based functional (LDA- or GGA-type),
\begin{equation} 
 E_c^{\mr{RPA+}} = E_\mr{c,sr}^{\mr{LDA}} + \EcRPA
\label{eq:RPA+}
\end{equation} 
where the LDA for the short-range contribution $E_\mr{c,sr}$ can be 
obtained by subtraction of the RPA-limit from the full LDA correlation 
energy,  
\begin{equation}  
 E_\mr{c,sr}^{\mr{LDA}} =  E_\mr{c}^{\mr{LDA}} - E_\mr{c}^{\mr{LDA-RPA}}
\label{eq:SRLDA}
\;.
\end{equation}
This approach is supported by the fact that the gradient correction to 
the short-range correlation is much smaller than that to the complete 
correlation energy.\cite{Kurth99} 
Though the RPA+ functional has been used recently to describe the 
inter-layer dispersion interaction in boron nitride,\cite{Marini06} a direct 
comparison of the RPA+ with exact results is still missing even for 
closed-shell atoms. 
Using numerically exact RPA correlation energy available for atoms, 
we are able to give a unambiguous assessment of the quality of the 
RPA+ correlation functional. 

In the language of Feynman diagrams, the RPA correlation energy is 
obtained from the second order direct diagram by replacing the 
bare Coulomb interaction by the dynamically screened Coulomb 
interaction. 
The dominant contribution that is missing in the RPA is the second 
order exchange diagram (SOX), 
\begin{equation} 
E_c^{\mathrm{SOX}} 
= 
- \frac{1}{2}\sum_{ij,ab} 
  \frac{ \TEI{ij}{ab} \TEI{ab}{ji} }{\eps_i+\eps_j-\eps_a-\eps_b}
\label{eq:SOX}
\;,
\end{equation} 
where the notation 
\begin{equation}
(pq||rs)
=
\int d^3r \int d^3r'\,
\frac{ \phi_p^\dagger(\bm{r} ) \phi_r(\bm{r} )
       \phi_q^\dagger(\bm{r}') \phi_s(\bm{r}')
}{
  \vert\bm{r} - \bm{r}'\vert
}
\end{equation}
has been used for the KS Slater integral. 
Combining the RPA with the SOX term, one obtains a new functional,
denoted as RPA+SOX. 
However, one would expect the SOX to over-correct the error in the
RPA, since the Coulomb interaction enters the SOX term in its bare,
i.e.\  un-screened, form. 
Screening can be introduced into the SOX term in a systematic way
by use of the same, dynamically screened interaction as in the direct
term. \cite{Hu86} 
Unfortunately, the resulting functional is computationally much more
demanding than the RPA expression.
A technically much simpler way to reduce the SOX contribution has been 
suggested in the context of the second order functional $E_c^{(2)}$:%
\cite{Jiang06,Engel06} 
The inclusion of the direct hole-hole contribution to the
Epstein-Nesbet-type ladder diagrams into the SOX term substantially 
improves second order energies and potentials, without introducing 
any additional computational effort.
Although the physical background of these ladder diagrams is quite 
different from dynamical screening, it seems worthwhile to analyze
this ``effective'' screening.
The resulting correction will be denoted as RSOX, 
\begin{equation} 
E_c^{\mathrm{RSOX}} 
= 
-\frac{1}{2}\sum_{i,j,a,b} 
\frac{ \TEI{ij}{ab} \TEI{ab}{ji} }
     {\eps_i+\eps_j-\eps_a-\eps_b - \TEI{ij}{ij} }
\label{eq:RSOX}
\;.
\end{equation}

\subsection{RPA correlation functional for spherical systems}
In the case of spherical systems, the KS potential only depends on the 
radial coordinate $r$, $\vs(\br) = \vs(r)$, and each KS orbital can be
written as the product of a radial orbital and a spherical harmonic 
$Y_{lm}(\theta,\varphi)$,
\begin{equation}
\phi_k(\br) \rightarrow \phi_{nlm}(\br)
=
\frac{P_{nl}(r)}{r}Y_{lm}(\theta,\varphi),
\;,
\end{equation}
where $n$, $l$ and $m$ are the principle, angular and magnetic quantum 
numbers, respectively.  
The $P_{nl}$ are solutions of the radial KS equation,
\begin{equation}
\left[ -\half\left(\frac{d^2}{dr^2}-\frac{l(l+1)}{r^2}\right)+\vs(r) \right] P_{nl}(r)
=
\eps_{nl}P_{nl}(r)
\label{eq:KS-eqs}
\;.
\end{equation}

Two-body functions like the Coulomb interaction and $\chi_0$ can also 
be decomposed according to the spherical symmetry. 
To simplify notations, we use the following decomposition convention. 
The Coulomb interaction is expanded as
\begin{eqnarray}
\vint(\br,\br') 
&=&
\sum_{L=0}^{\infty} \frac{4\pi}{2L+1} v_L(r,r') 
\nonumber
\\
&&\times
\sum_{M=-L}^L Y_{LM}(\theta,\varphi) Y_{LM}^*(\theta',\varphi')
\label{eq:v-decomp}
\;,
\end{eqnarray}
where 
\begin{equation}
 v_L(r,r') \Def r_<^L / r_>^{L+1}
\end{equation}
with $r_< = \mathrm{Min}(r,r')$ and $r_> = \mathrm{Max}(r,r')$.
The response function $\chi_0$ can be written as 
\begin{align}
\chi_0(\br,\br',iu) 
=& 
\sum_{L=0}^{\infty} 
\frac{2L+1}{4\pi} 
\frac{\chi_{0L}(r,r',iu)} {r^2 r'^2} 
\nonumber 
\\
&\times  
\sum_{M=-L}^L Y_{LM}(\theta,\varphi) Y_{LM}^*(\theta',\varphi')
\label{eq:chi0-decomp}
\;.
\end{align}
The $L$-dependent radial response function $\chi_{0L}(r,r',iu)$ can be 
calculated utilizing the radial orbitals 
\begin{equation}
\chi_{0L}(r,r',iu) 
=
- \sum_{ia\sigma} 
  C_{L;ia\sigma} D_{ia\sigma}(u) \Phi_{ia\sigma}(r) \Phi_{ia\sigma}(r')
\label{eq:rad-chi0}
\,,
\end{equation}
where 
\begin{eqnarray}
C_{L;ia\sigma} 
&\Def& 
\frac{(2l_i+1)(2l_a+1)}{2L+1} \ThreeJ{l_i}{l_a}{L}{0}{0}{0}^2   \label{eq:CLia}
\\
D_{ia\sigma}(u) 
&\Def& 
\frac{2(\eps_{a\sigma} - \eps_{i\sigma})}{u^2 + (\eps_{a\sigma} - \eps_{i\sigma} )^2}  \label{eq:Dia}
\\
\Phi_{ia\sigma}(r) 
&\Def&  
P_{i\sigma} (r)P_{a\sigma}(r) 
\;.\label{eq:Pia}
\end{eqnarray}

Using the multipole expansion of both $\vint$ and $\chi_0$, the 
building block of the RPA correlation energy, $\Tr{\chi_0 v}$, 
is be obtained as 
\begin{equation}
\Tr{\chi_0 v} 
= 
\sum_{L} (2L+1) \int_0^\infty dr dr' \chi_{0L}(r,r',iu) v_L(r',r) 
\label{eq:chi0v}
\;.
\end{equation}
There are two options for the calculation of the radial integral in 
Eq. (\ref{eq:chi0v}):

\paragraph{Real space approach:} 
In this approach, $\chi_{0L}(r,r')$ is calculated on a discrete 
radial mesh, which allows to evaluate (\ref{eq:chi0v}) by direct 
numerical integration, 
\begin{eqnarray}
&&
\Tr{\chi_0 v} 
\nonumber \\
&=& 
\sum_{L}(2L+1) \sum_{i,j} 
w(r_i) w(r_j) \chi_{0L}(r_i,r_j)  v_L(r_j,r_i) 
\nonumber \\
&=& 
\sum_{L}(2L+1) \sum_{i,j} 
\left[\tilde{\chi}_{0L}\right] _{i,j} \left[v_{L}\right]_{j,i} 
\nonumber \\
&=&
\sum_{L}(2L+1) \Tr{ \tilde{\chi}_{0L} v_L }
\;,
\end{eqnarray}
where $w_i$ denotes the radial integral weight at mesh point $i$. 
In case of powers of $\Tr{\chi_0 v}$ one has 
\begin{equation}
\Tr{(\chi_0 v)^n} 
=  
\sum_{L}(2L+1) \Tr{ (\tilde{\chi}_{0L} v_L)^n }
\;.
\end{equation}
The sum over $n$ in Eq. (\ref{eq:EcRPA-2}) then leads to 
\begin{equation}
E_c(iu) 
=  
\sum_{L}(2L+1) 
\Tr{ \Ln{ 1-\tilde{\chi}_{0L} v_L } + \tilde{\chi}_{0L} v_L}
\;.
\end{equation} 

\paragraph{Orbital-product space approach:}
In this second approach one inserts (\ref{eq:chi0-decomp}) and
(\ref{eq:rad-chi0}) into Eq. (\ref{eq:chi0v}). 
Using the radial Slater integrals
\begin{equation}
R_{L;ia\sigma,jb\sigma'} 
\Def 
\int_0^\infty dr \int_0^\infty dr' \Phi_{ia\sigma}(r) v_L(r,r')\Phi_{jb\sigma'}(r') 
\label{eq:Slater-int}
\;,
\end{equation}
one finds
\begin{eqnarray}
\Tr{\chi_0 v}
&=& 
\sum_{L} (2L+1) \sum_{ia\sigma}  C_{L;ia\sigma}  D_{ia\sigma}(u) 
\nonumber \\
& &\times   
\int dr \int dr' \Phi_{ia\sigma}(r) v_L(r,r')\Phi_{ia\sigma}(r')  
\nonumber \\
&=&  
- \sum_{L} (2L+1) \sum_{ia\sigma}
  C_{L;ia\sigma}  D_{ia\sigma}(u)  R_{L;ia\sigma,ia\sigma}
\nonumber \\
\end{eqnarray}
With the definitions
\begin{eqnarray}
V_{L;ia\sigma,jb\sigma'} 
&\Def& 
\sqrt{C_{L;ia\sigma}} R_{L;ia\sigma,jb\sigma'} \sqrt{C_{L;jb\sigma'}} 
\label{eq:VL}
\\
S_{L;ia\sigma,jb\sigma'} 
&\Def&  
-
\sqrt{D_{ia\sigma}(u)} V_{L;ia\sigma,jb\sigma'} \sqrt{D_{jb\sigma'}(u)}
\phantom{shi}
\end{eqnarray}
one ends up with
\begin{eqnarray}
\Tr{\chi_0 v} 
&=&  
\sum_{L} (2L+1) \sum_{ia\sigma} S_{L;ia\sigma,ia\sigma} 
\nonumber \\ 
&=&  
\sum_{L} (2L+1) \Tr{S_L}
\;.
\end{eqnarray}
One can furthermore show that
\begin{equation}
\Tr{(\chi_0 v)^n} = \sum_{L} (2L+1)  \Tr{(S_L)^n}
\end{equation} 
so that 
\begin{eqnarray}
E_c(iu) 
&=& 
- \sum_{L} (2L+1) \Tr{ \sum_{n=2}^\infty \frac{(S_L)^n }{n}  } 
\nonumber \\
&=& 
\sum_{L} (2L+1)\Tr{ \Ln{1-S_L} + S_L} 
\nonumber \\
&=& 
\sum_{L} (2L+1) \lbrack \Ln{\Det{1-S_L}} + \Tr{S_L} \rbrack 
\label{eq:Ecu-3}
\;. 
\phantom{sh}
\end{eqnarray} 

The final expressions for $\EcRPA$ are quite similar in the real-space 
and orbital-product-space approaches, but their numerical efficiency 
can be very different, depending on the size of the system. 
In the real-space approach, the dimension of the matrix involved is 
determined by the number of mesh points used for radial integration,
$\Imax$.
As $\Imax$ is never larger than a few thousand even for heavy atoms, 
the resulting memory requirement is quite low. 
On the other hand, $\chi_{0L}$ needs to be constructed on the radial 
mesh for each frequency, which can be very cpu-time-intensive. 
The situation is quite different in the case of the orbital-product 
space approach.
Here the dimension of the matrix involved is given by 
$N_{\rm occ} \times N_{\rm vir}$, where $N_{\rm occ}$ denotes the
number of occupied orbitals and $N_{\rm vir}$ is the number of 
unoccupied orbitals taken into account. 
$N_{\rm occ} \times N_{\rm vir}$ can be easily as high as tens of 
thousands. 
However, the matrix $V_L$, Eq.(\ref{eq:VL}), is independent of 
frequency and can be calculated in advance and stored in the memory.
Limitations of the available memory can be circumvented by taking 
advantage of the fact that, for given $L$, $V_L$ and $S_L$ are quite 
sparse (usually the ratio of non-zero elements is less than $1\%$). 
One can therefore use standard sparse matrix techniques to reduce 
the storage requirement and accelerate the computation of the 
determinant.

\section{Numerical details}
\subsection{Hard-wall cavity approach}
The RPA correlation energy depends on the occupied as well as on the 
unoccupied KS states. 
For free atoms, the spectrum of the unoccupied states includes both 
discrete Rydberg states and continuum states. 
However, the handling of continuum states in the evaluation of the
correlation energy is nontrivial. \cite{Kelly64}.
Moreover, the presence of continuum states causes additional problems 
in the context of orbital-dependent functionals:
One does no longer find a solution of the corresponding OPM equation 
which satisfies the standard boundary conditions for the correlation
potential.\cite{Engel05}  
To resolve this problem, we developed a hard-wall cavity approach, 
\cite{Engel05,Jiang05, Engel06} in which the KS equation is solved 
on a discrete radial mesh with hard-wall boundary conditions imposed
at a finite, but large radius $\Rmax$. 
The same approach is used in the present work. 

Its crucial parameters are the cavity radius $\Rmax$ as well as the 
energetically highest state (characterized by its principle quantum 
number $\nmax$) and the highest angular momentum $\lmax$ included in 
sums over virtual states. 
In the following, neutral Ar is used for a systematic convergence study 
with respect to $\Rmax$, $\nmax$ and $\lmax$. 

\begin{table}
\caption{\label{tab:conv-Rmax} 
Convergence of $\EcRPA$, the exact exchange energy and the eigenvalue
of the highest occupied KS orbital ($\eps_{\rm HOMO}$) obtained by 
self-consistent exchange-only calculations for Ar for different
cavity radii $\Rmax$ (with $\nmax/\Rmax=10$ Bohr$^{-1}$ and $\lmax=4$,
$\Rmax$ in Bohr, all energies in Hartree).}
\begin{center}
\begin{tabular}{cccc}
\hline 
$\Rmax$ & $-\EcRPA$   & $-E_x$  & $\eps_{\rm HOMO}$ \\
\hline 
5        &   1.0023     & 30.2059  & 0.5772 \\
8        &   1.0027     & 30.1749  & 0.5909 \\
10       &   1.0028     & 30.1747  & 0.5908 \\
\hline 
\end{tabular}
\end{center}
\end{table}

\begin{table}
\caption{\label{tab:conv-nmax} 
Convergence of full $\EcRPA$ (Column 3) and $\EcRPA$ within the frozen 
core approximation excluding virtual excitations of the $1s$, $2s$ and 
$2p$ electrons (Column 4) of Ar with respect to $\nmax$ 
(with $\Rmax=10$ Bohr and $\lmax=4$, all energies in Hartree).   }
\begin{center}
\begin{tabular}{cccc}
\hline 
$\nmax$& $\emax$ & $-\EcRPA$    & $-\EcRPA$(FC) \\
\hline 
  25   & 25.1    &   0.6840      &  0.3961     \\           
  50   & 111.9   &   0.9097      &  0.3980     \\
 100   & 471.2   &   1.0028      &  0.3980    \\
 150   & 1077.6  &   1.0273      &  0.3980     \\
 200   & 1930.8  &   1.0354      &  0.3980     \\
 250   & 3030.9  &   1.0384      &  0.3980     \\
 300   & 4377.6  &   1.0398      &  0.3980     \\
 350   & 5951.9  &   1.0399      &            \\     
 400   & 7754.8  &   1.0400      &            \\
\hline 
\end{tabular}
\end{center}
\end{table}

\begin{table}
\caption{\label{tab:conv-lmax} 
Convergence of full $\EcRPA$ (Column 3) and $\EcRPA$ within the frozen 
core approximation excluding virtual excitations of the $1s$, $2s$ and 
$2p$ electrons (Column 4) of Ar with respect to $\lmax$ 
(with $\Rmax=10$ Bohr and $\nmax=100$, all energies in Hartree). }
\begin{center}
\begin{tabular}{ccc}
\hline 
 $~\lmax$ &~$-\EcRPA$ & $-\EcRPA$(FC)   \\
\hline 
  2       & 0.7661     & 0.2875     \\
  4       & 1.0028     & 0.3980     \\
  6       & 1.0431     & 0.4185     \\
  8       & 1.0529     & 0.4241     \\
 10       & 1.0562     & 0.4262     \\
 12       & 1.0574     & 0.4271     \\
 14       & 1.0580     &            \\
 16       & 1.0582     &            \\          
\hline 
\end{tabular}
\end{center}
\end{table}

We first consider $\Rmax$. 
$\Rmax$ has to be chosen so large, that all ground state properties, 
and, in particular, the correlation energy, do no longer change when 
$\Rmax$ is increased further. 
However, any increase of $\Rmax$ directly affects the spectrum of the
positive energy states, i.e.\  the density of states.
In order to keep the space available for virtual excitations constant, 
when increasing $\Rmax$, one therefore has to fix the energy $\emax$
of the highest unoccupied state $\nmax$ taken into account.
In the case of very high-lying virtual states one has a simple relation
between $\Rmax$ and $\emax$, resulting from the fact that high-lying 
states are no longer sensitive to the detailed structure of $\vs$,
\begin{equation}
 \emax \propto \left( \frac{\nmax}{\Rmax} \right) ^2
\;.
\end{equation}
The space available for virtual excitations is therefore kept constant,
as soon as the ratio $\nmax/\Rmax$ is fixed.
Table \ref{tab:conv-Rmax} shows the values of $\EcRPA$ for Ar obtained 
with different $\Rmax$, but fixed $\nmax/\Rmax=10\,$Bohr$^{-1}$, which
corresponds to an energy cut-off of about 500 Hartree. 
For comparison, the corresponding exchange energy and the eigenvalue 
of the highest occupied orbital, $\eps_{\rm HOMO}$, resulting from 
exchange-only calculations are also listed. 
One observes that $\EcRPA$ is less sensitive to $\Rmax$ than the 
exchange energy, which is consistent with the fact that the length 
scale related to the RPA correlation energy is smaller compared to 
that of the exchange. 

Argon is the heaviest atom considered in this work.
We have also made systematic convergence tests for other, less compact 
atoms like Na and Mg. 
For all atoms considered in this work, the choice $\Rmax=10$ Bohr leads
to errors less than 1\,mHartree. 

With $\Rmax$ fixed, one can now examine the convergence of $\EcRPA$
with respect to $\nmax$ and $\lmax$.
Tables \ref{tab:conv-nmax} and \ref{tab:conv-lmax} show $\EcRPA$ for Ar 
obtained with different $\nmax$ and $\lmax$. 
In general, the absolute value of $\EcRPA$ converges quite slowly with 
respect to both parameters.
The slow convergence with respect to $\nmax$ mainly originates from 
the innermost shell --- unoccupied states with high energies are only
important for the description of virtual excitations of the highly
localized, low-lying core states.
In practice, fortunately only energy differences related to the valence
electrons are really relevant.
One would thus expect to achieve high accuracy for these energy 
differences with much more moderate values for $\nmax$.
This suggests to rely on the frozen core (FC) approximation, in which
excitations from core levels are excluded. 
Tables \ref{tab:conv-nmax} and \ref{tab:conv-lmax} demonstrate that
the FC approximation for $\EcRPA$ converges much faster with increasing 
$\nmax$. 
Even for a quite moderate $\nmax$ of 25, corresponding to 
$\emax \sim 25$\,Hartree, $\EcRPA$ is already converged to an accuracy 
of 2 mHartree. 
On the other hand, the convergence behavior of $\EcRPA$ with respect 
to $\lmax$ is not improved by the FC approximation. 
As one of the main aims of this work is to provide benchmark results 
for a set of prototype atoms, most results reported in this work are 
obtained without evoking the FC approximation. 
The results reported in the next section are obtained for $\nmax=300$ 
and $\lmax=14$, which ensures an accuracy of 1\,mHartree for Ar and 
better for all lighter atoms.


\subsection{Frequency integration}

\begin{figure}
\centerline{\includegraphics[width=2.8in,clip,angle=-90]{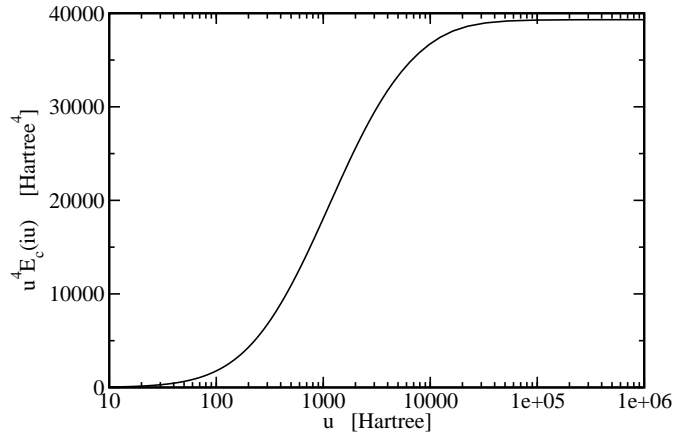}}
\caption{\label{fig:u4Ec-u-He} (Color online)
$u^4E_c(iu)$ vs $u$ for He 
(maximum excitation energy $\delta\epsilon\approx4000$\,Hartree).
}
\end{figure}

Any calculation of RPA energies involves two time-consuming steps:
The first is the evaluation of all Slater integrals involved, i.e.\
of the matrix $R_L$, Eq.(\ref{eq:Slater-int}).
In the present work the Slater integrals are calculated by numerical
integration on the radial grid, using standard finite differences 
methods.
Once $R_L$ is available, the second step is performing the frequency 
integration in Eq.(\ref{eq:EcRPA}).
In order to understand the most appropriate way to do this frequency
integration let us consider the integrand for some prototype atoms.
Figure \ref{fig:u4Ec-u-He} shows $u^4E_c(iu)$ for He, demonstrating the
fact that $E_c(iu)$ falls off as $u^{-4}$ for extremely large $u$.
This behavior can be easily understood on the basis of Eq. (\ref{eq:Dia}):
For frequencies beyond the maximum excitation energy 
$\delta\epsilon=\epsilon_{a\sigma}^{\rm max}-\epsilon_{i\sigma}$ 
included in the calculation (or provided by the basis set) 
$D_{ia\sigma}(u)$ and thus $S_L(u)$ decay as $u^{-2}$ which allows 
a perturbative evaluation of (\ref{eq:Ecu-3}) in powers of $S_L(u)$, 
with the second order term dominating the resulting $E_c(iu)$.

\begin{figure}
\centerline{\includegraphics[width=2.8in,clip,angle=-90]{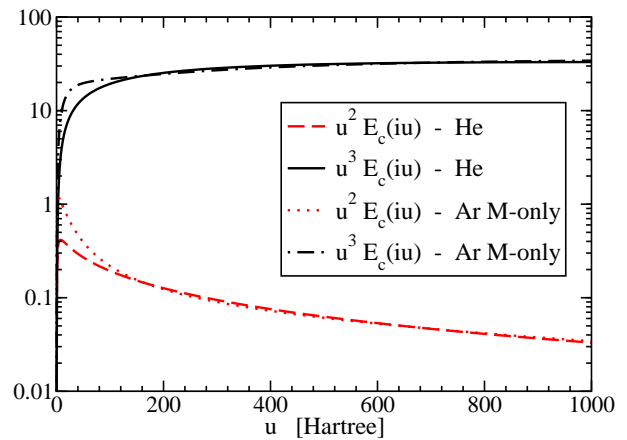}}
\caption{\label{fig:u23Ec-u-He} (Color online)
$u^2E_c(iu)$ and $u^3E_c(iu)$ vs $u$ for moderately large $u$
for the case of He.
}
\end{figure}

On the other hand, for the more important range of large frequencies 
less than $\delta\epsilon$ a decay close to $u^{-3}$ is found, as 
shown in Figure \ref{fig:u23Ec-u-He}.
The same behavior is observed for each individual shell, as illustrated 
by the $E_c(iu)$ obtained by excitation of only the $M$-shell of neutral
Ar, also included in Figure \ref{fig:u23Ec-u-He}.

\begin{figure}
\centerline{\includegraphics[width=2.8in,clip,angle=-90]{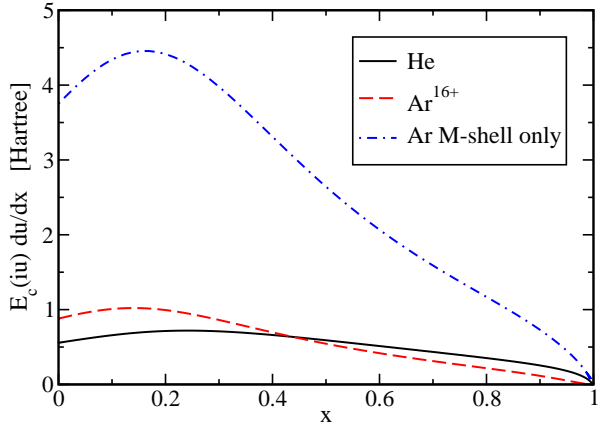}}
\caption{\label{fig:trf-single-shell} (Color online)
Integrand $E_c(iu)du/dx$ of frequency integral after the transformation
(\protect\ref{trf-single-shell}) vs $x$ for He,  Ar$^{16+}$ as well as
the $M$-shell of neutral Ar.
}
\end{figure}

This power law decay suggests a transformation of the frequency interval
$0\le u<\infty$ to some finite interval via a power law transformation,
as for instance
\begin{equation}
x=\frac{1}{1+(u/s)}
\hskip30pt
0 \le x \le 1
\label{trf-single-shell}
\;,
\end{equation}
giving more weight to large $u$ than an exponential transformation.
The scale factor $s$ is intrinsically related to the minimum energy
required for a virtual excitation, which is roughly given by the 
eigenvalue difference $\epsilon_{LUMO}-\langle\epsilon\rangle
\approx|\langle\epsilon\rangle|$ for a shell with average eigenvalue
$\langle\epsilon\rangle$.
The most appropriate $s$ can only be determined empirically.
For all atoms considered in detail the choice
$s=2|\langle\epsilon\rangle|$ seemed to work reasonably well 
(see also below).
The result of the transformation (\ref{trf-single-shell}) is shown
in Figure \ref{fig:trf-single-shell} for He, Ar$^{16+}$ as well as
the $M$-shell of neutral Ar. 
One obtains a smooth function of $x$, with values remaining on the
same order of magnitude for all $x$.
This ensures a rapid convergence of the numerical integration over $x$
with the number of grid points.

\begin{figure}
\centerline{\includegraphics[width=2.8in,clip,angle=-90]{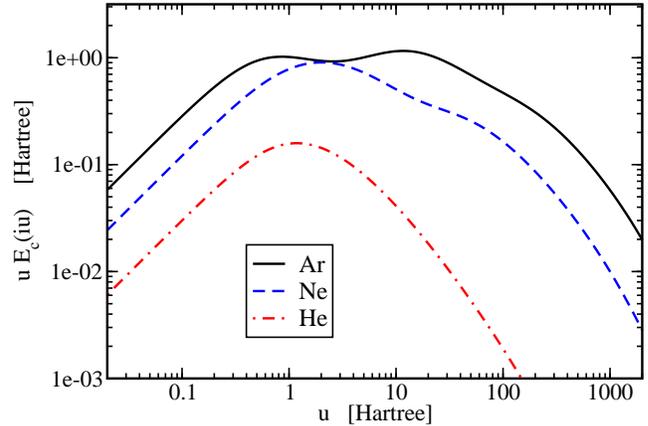}}
\caption{\label{fig:Ec-u-Ne-Ar} (Color online)
$uE_c(iu)$ vs $u$ for Ne and Ar.
}
\end{figure}

\begin{figure}
\centerline{\includegraphics[width=2.8in,clip,angle=-90]{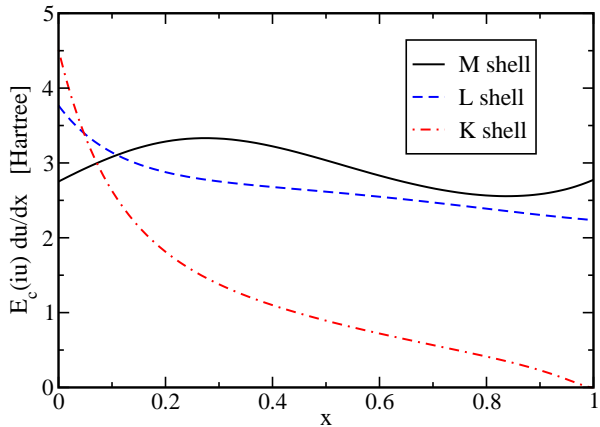}}
\caption{\label{fig:trf-all-shells} (Color online)
Integrand $E_c(iu(x))du/dx$ of partioned frequency integral
(\protect\ref{fre-int-split-4}) resulting from the transformation 
(\protect\ref{fre-int-split-2}) for neutral Ar.
}
\end{figure}

\begin{figure}
\centerline{\includegraphics[width=2.8in,clip,angle=-90]{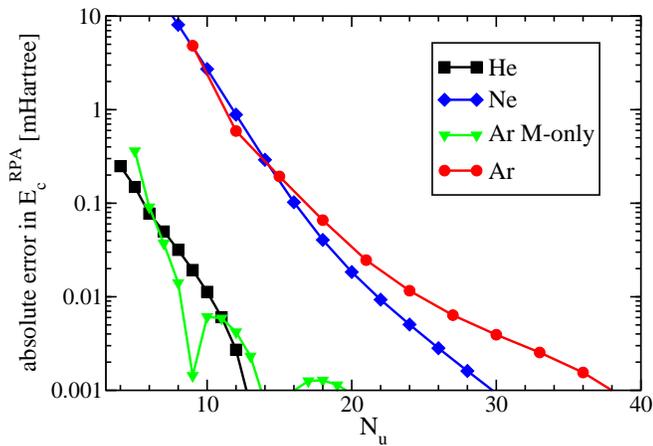}}
\caption{\label{fig:gauleg-conv} (Color online)
Absolute error resulting from Eqs.(\protect\ref{fre-int-split-4}) and
(\protect\ref{fre-int-split-2}) as a function the total number of grid 
points $N_u$ used for the Gauss-Legendre quadrature (note: $N_u$ is
the sum of the number of grid points used in the individual intervals).
}
\end{figure}

However, the frequency integration in (\ref{eq:EcRPA}) suffers from 
the fact that each shell in an atom (or molecule) introduces a new 
energy scale for the virtual excitations.
This is most easily verified by plotting $uE_c(iu)$ on a double
logarithmic scale, as done in Figure \ref{fig:Ec-u-Ne-Ar} for He,
Ne and Ar.
Figure \ref{fig:Ec-u-Ne-Ar} demonstrates that there are two relevant 
scales for Ne, three in the case of Ar.
In fact, the plots confirm that the behavior of $E_c(iu)$ changes 
at roughly twice the average eigenvalues of the shells involved,
with the position of the highest energy transition point being 
somewhat less pronounced
($\epsilon_K$(Ne)=$-30.8$, $\epsilon_L$(Ne)=$-1.1$;
$\epsilon_K$(Ar)=$-114.4$, $\epsilon_L$(Ar)=$-9.4$,
$\epsilon_M$(Ar)=$-0.7$ --- all values in Hartree).
It is therefore necessary to split the frequency integration from 0 to 
$\infty$ into intervals associated with these individual energy scales.
Let us call the boundaries of the intervals $b_i$,
\begin{equation}
0 = b_0 < b_1 < \ldots < b_n = \infty
\end{equation}
with $n$ denoting the number of shells.
The intervals are chosen such that the characteristic energy scale
$s_n$ of shell $n$ is bracketed,
\begin{equation}
b_{i-1} < s_i < b_i
\;.
\end{equation}
In practice, $s_i=2|\langle\epsilon_i\rangle|$ and
$b_i=4|\langle\epsilon_i\rangle|$ seemed to provide a reasonable
partioning of the complete frequency range.
Eq.(\ref{eq:EcRPA}) may then be decomposed as
\begin{eqnarray}
\int_{0}^{\infty} E_c(iu) du
=
\sum_{i=1}^n
\int_{b_{i-1}}^{b_i} E_c(iu) du
\label{fre-int-split}
\;.
\end{eqnarray}
In order to account for the piecewise decay of $E_c(iu)$ the 
transformation
\begin{eqnarray}
x_i
&=&
\frac{\lbrack 1+s_i/(b_i-b_{i-1}) \rbrack}
     {\lbrack 1+(u  -b_{i-1})/s_i \rbrack}\,
\frac{u-b_{i-1}}{s_i}
\label{fre-int-split-2}
\\
\frac{du}{dx_i}
&=&
s_i\,
\frac{\lbrack 1+s_i/(b_i-b_{i-1}) \rbrack}
     {\lbrack 1+s_i/(b_i-b_{i-1})-x_i \rbrack^2}
\label{fre-int-split-3}
\end{eqnarray}
$(i=1,\ldots n)$ is most suitable.
Eq.(\ref{fre-int-split}) can then be written as
\begin{eqnarray}
\int_{0}^{\infty} E_c(iu) du
&=&
\sum_{i=1}^n
\int_{-1}^{1} dx_i \frac{du}{dx_i} E_c(iu(x_i))
\label{fre-int-split-4}
\;.
\end{eqnarray}
The success of this frequency partioning plus power law transformation 
scheme is demonstrated in Figure \ref{fig:trf-all-shells}, in which the 
final integrands $du/dx_i\,E_c(iu(x_i))$ are plotted for neutral Ar.
In all three energy regimes a rather smooth integrand is obtained,
which allows the application of Gauss-Legendre quadrature to each
interval.
As a result, the error obtained for a given total number of grid points 
$N_u$ used for the Gauss-Legendre quadrature is rather small already 
for very low $N_u$, as illustrated in Figure \ref{fig:gauleg-conv}.
The most critical interval in Eq.(\ref{fre-int-split-4}) is the highest 
energy range, covering in particular excitations of the $1s$-state.
For that reason the error is even lower if only excitations of the 
valence shell are included (i.e.\  in the FC approximation), as is
obvious from the error found for He or the $M$-shell of neutral Ar.

\section{Results}

\subsection{Sensitivity to form of KS orbitals}

\begin{table*}
\caption{\label{tab:input} 
Absolute RPA total energies (in Hartree) resulting from insertion of 
different KS orbitals. 
The last row lists the self-consistent RPA total energies given in 
Ref.\onlinecite{Hellgren07}.}
\begin{tabular}{c|ccccccc}
\hline
\hline 
KS orbitals&  He     &  Be      &  Ne       &   Mg      & Ar        &  N       &    Na   \\
\hline
LDA        &~~2.945~~&~~14.751~~&~~129.140~~&~~200.293~~&~~527.905~~&~~54.735~~&~~162.475~~\\
BLYP       &~~2.944~~&~~14.752~~&~~129.142~~&~~200.296~~&~~527.910~~&~~54.737~~&~~162.478~~\\
EXX-only   &~~2.945~~&~~14.752~~&~~129.143~~&~~200.298~~&~~527.913~~&~~54.738~~&~~162.480~~\\
RPA        &~~2.945~~&~~14.754~~&~~129.143~~&~~200.296~~&~~527.908~~&   ---   &   ---  \\
\hline 
\hline
\end{tabular}
\end{table*}

\begin{figure}
\centerline{\includegraphics[width=3in,clip]{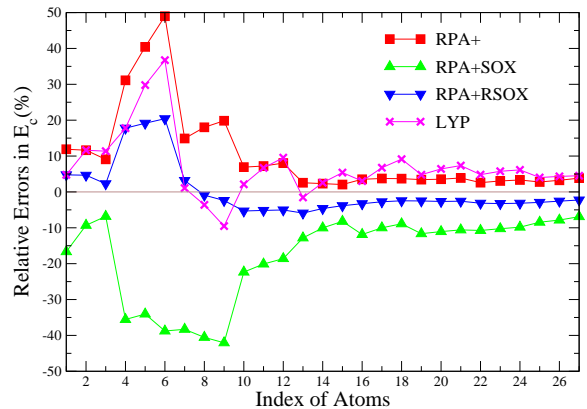}}
\caption{\label{fig:Ec} (Color online)
Relative errors resulting from different approximate correlation 
energies, obtained from the data in Table \protect\ref{tab:Ec}.
}
\end{figure}

Standard KS-DFT calculations are based on the self-consistent solution 
of the KS equations, which requires the evaluation of the XC potential 
$\vxc(\br)=\delta\Exc[n]/\delta n(\br)$. 
In the case of LDA and GGA functionals the calculation of $\vxc(\br)$ 
is straightforward. 
On the other hand, a self-consistent implementation represents a much
more serious problem for RPA-type functionals. 
First of all, in the case of orbital-dependent functionals $\vxc$ has 
to be determined via the OPM, i.e.\  by solution of an integral 
equation.\cite{Talman76} 
The solution of the OPM integral equation is well-established for the
exact exchange and managable, though rather intricate, for the second 
order correlation functional $E_c^{(2)}$.\cite{Grabowski02,Yang02,%
MoriSanchez05,Jiang05}
Its implementation for RPA-type functionals, however, is much more 
challenging, so that a full solution has only been reported very 
recently.\cite{Hellgren07} 
On the other hand, a self-consistent implementation is only then
advantageous if the resulting correlation potential leads to an 
improved total KS potential.
It has been demonstrated that this is not the case for some standard 
GGAs \cite{Umrigar94} and for $E_c^{(2)}$, \cite{Jiang05,Jiang06} 
which by far overestimates correlation effects.
The situation is not yet completely clear in the case of the RPA.
The first RPA potentials available \cite{Hellgren07} seem to improve
over $v_c^{(2)}=\delta E_c^{(2)}/\delta n(\br)$ in the asymptotic 
regime, but otherwise often follow $v_c^{(2)}$.

Independently of this more fundamental aspect, one might ask whether
a self-consistent implementation is really necessary to obtain accurate 
RPA correlation and thus ground state energies.
Clearly, a purely perturbative treatment of RPA-type functionals on 
the basis of a self-consistent calculation with the exact exchange
would allow their application to much more complex systems, for which
the solution of the corresponding OPM integral equation is beyond
current computer resources.
In fact, experience with conventional density functionals suggests that,
at least for atomic systems, the RPA correlation energy is not sensitive
to the detailed structure of $\vxc(\br)$.

In order to verify this expectation, the RPA ground state energy
(i.e.\ the sum of the KS kinetic energy, the Hartree term, the exact 
exchange and $\EcRPA$) has been calculated by insertion of KS orbitals 
resulting from different XC functionals:
Orbitals obtained from self-consistent calculations with only the exact 
exchange (EXX-only), but neglecting $\vc$ completely, are compared with 
self-consistent LDA and GGA orbitals.
The results for a number of atoms are collected in Table 
\ref{tab:input}, which also includes recent self-consistent RPA 
energies,\cite{Hellgren07} whenever available. 
Table \ref{tab:input} confirms the expectation: 
Even though the KS potentials obtained by EXX-only calculations
differ substantially from their LDA and GGA counterparts, the 
differences between the resulting RPA energies are quite small. 
The same is true for the deviations between the perturbative RPA
energies on EXX-only basis and the self-consistent RPA results.
This result is expected to hold quite generally, as long as one 
does not examine a quantity which is particularly sensitive to the 
correlation potential (as, for instance, the quantum defect of high 
Rydberg states \cite{Faassen06}).
In fact, Table \ref{tab:input} indicates that even a perturbative 
treatment of both the exact exchange and the RPA may be legitimate 
for very complex systems, in which even self-consistent calculations 
with the exact exchange are too expensive.
In the following sections, self-consistent EXX-only orbitals are
always used for the evaluation of the RPA correlation energy.

\subsection{RPA correlation energies of spherical atoms}
In this work, we focus on atoms with closed or half-filled shells, 
for which the ground state KS potentials for the two spin-channels 
are both spherically symmetric.
\begin{table}
\caption{\label{tab:Ec}
Absolute correlation energies (in Hartree) of closed sub-shell atoms 
calculated from the RPA and RPA+ functionals by insertion of the exact 
EXX-only KS orbitals in comparison with exact data.
\cite{Chakravorty93}
The last row provides the mean absolute error (MAE) with respect to the exact energies.}
\begin{tabular}{lccccccc}
\hline
\hline
atom     &~~exact~~&~~RPA~~  &~~RPA+~~& RPA+SOX & RPA+RSOX& LYP  \\
\\
\hline
He	 &  0.042  &  0.083  & 0.047  &  0.035  &  0.044  & 0.044 \\ 
Li$^+$   &  0.043  &  0.087  & 0.048  &  0.039  &  0.045  & 0.048 \\
Be$^{2+}$&  0.044  &  0.088  & 0.048  &  0.041  &  0.045  & 0.049 \\ 
Li	 &  0.045  &  0.112  & 0.059  &  0.029  &  0.053  & 0.053 \\
Be$^+$   &  0.047  &  0.122  & 0.066  &  0.031  &  0.056  & 0.061 \\
B$^{2+}$ &  0.049  &  0.131  & 0.073  &  0.030  &  0.059  & 0.067 \\
Be       &  0.094  &  0.179  & 0.108  &  0.058  &  0.097  & 0.095 \\
B$^+$    &  0.111  &  0.205  & 0.131  &  0.066  &  0.110  & 0.107 \\
C$^{2+}$ &  0.126  &  0.228  & 0.151  &  0.073  &  0.123  & 0.114 \\
N	 &  0.188  &  0.335  & 0.201  &  0.146  &  0.178  & 0.192 \\
O$^+$    &  0.194  &  0.345  & 0.208  &  0.155  &  0.184  & 0.207 \\
F$^{2+}$ &  0.199  &  0.355  & 0.215  &  0.162  &  0.189  & 0.218 \\
Ne	 &  0.390  &  0.597  & 0.400  &  0.340  &  0.367  & 0.384 \\ 
Na$^+$   &  0.389  &  0.599  & 0.398  &  0.350  &  0.371  & 0.399 \\ 
Mg$^{2+}$&  0.390  &  0.601  & 0.398  &  0.358  &  0.375  & 0.411 \\
Na	 &  0.396  &  0.626  & 0.410  &  0.349  &  0.383  & 0.408 \\
Mg$^+$   &  0.400  &  0.634  & 0.415  &  0.360  &  0.389  & 0.427 \\
Al$^{2+}$&  0.405  &  0.642  & 0.420  &  0.369  &  0.395  & 0.442 \\
Mg	 &  0.438  &  0.687  & 0.453  &  0.387  &  0.427  & 0.459 \\
Al$^+$   &  0.452  &  0.706  & 0.468  &  0.402  &  0.440  & 0.481 \\
Si$^{2+}$&  0.463  &  0.722  & 0.481  &  0.414  &  0.451  & 0.497 \\
P	 &  0.540  &  0.850  & 0.554  &  0.482  &  0.523  & 0.566 \\ 
S$^+$    &  0.556  &  0.873  & 0.573  &  0.499  &  0.538  & 0.588 \\
Cl$^{2+}$&  0.570  &  0.893  & 0.589  &  0.514  &  0.552  & 0.605 \\
Ar	 &  0.722  &  1.101  & 0.742  &  0.661  &  0.701  & 0.751 \\
K$^+$    &  0.739  &  1.126  & 0.763  &  0.681  &  0.720  & 0.771 \\
Ca$^{2+}$&  0.754  &  1.150  & 0.783  &  0.702  &  0.737  & 0.788 \\
\hline
MAE      &         &  0.196  & 0.015  &  0.039  &  0.011  & 0.018 \\
\hline
\hline
\end{tabular}
\end{table}
Table \ref{tab:Ec} lists the correlation energies obtained from all
four RPA-based functionals for a series of atoms.
To see trends more clearly, the relative errors resulting from the
various functionals with respect to the exact correlation energies 
\cite{Chakravorty93} are plotted in Figure \ref{fig:Ec}. 
Not surprisingly, the pure RPA always overestimates the true 
correlation energy.
Adding the short-range correction within the LDA (RPA+) improves the 
results remarkably, reducing the mean absolute error by more than an 
order of magnitude.
As expected, the unscreened SOX contribution by far overcorrects the 
error of the RPA. 
On the other hand, the inclusion of EN-corrections into the SOX term 
(RSOX) reduces this overcorrection significantly. 
In general, both the RPA+ and the RPA+RSOX produce more accurate 
correlation energies than the LYP-GGA, at least for the set of 
atoms considered in this work.  
Moreover, for light atoms one observes a tendency of the RPA+RSOX
to be superior to the RPA+.

\subsection{Ionization potentials}
Even more important than the accuracy of total (correlation) energies
is the accuracy of energy differences.
In the case of atoms the ionization potential (IP) serves as the 
prototype energy difference for assessing the quality of any 
approximation.
Much more than the total atomic $E_c$, the IP probes the description 
of the correlation of the valence states.
We have therefore calculated the IPs resulting from the four RPA-based 
correlation functionals for a number of atoms.
In order to avoid any uncertainty associated with spherical averaging, 
the comparison is restricted to atoms for which both the KS potential 
of the neutral ground state and that corresponding to the ionic state 
are spherical.
\begin{table*}
\caption{\label{tab:IP}
First ionization potentials (in Hartree) of spherical atoms calculated 
from total energy differences (IP$=E_{\rm tot}(N-1))-E_{\rm tot}(N)$),
using different XC functionals. 
The last row provides the mean absolute error (MAE) with respect to the 
exact results.\cite{Chakravorty93} 
Self-consistent EXX-only KS orbitals are used as input orbitals.}
\begin{tabular}{lccccccc}
\hline
\hline
atom     &~~exact~~&~~EXX~~ &~~RPA~~ &~~RPA+~~& RPA+SOX & RPA+RSOX &~~BLYP~~  \\
\hline 
Li     & 0.198 & 0.195 & 0.220 & 0.205  & 0.185 &  0.203 &  0.201 \\
Be$^+$ & 0.669 & 0.666 & 0.700 & 0.683  & 0.656 &  0.677 &  0.681 \\
Be     & 0.343 & 0.295 & 0.352 & 0.338  & 0.323 &  0.336 &  0.329 \\
B$^+$  & 0.924 & 0.861 & 0.935 & 0.919  & 0.897 &  0.912 &  0.904 \\
Na     & 0.189 & 0.179 & 0.206 & 0.191  & 0.178 &  0.191 &  0.194 \\
Mg$^+$ & 0.552 & 0.540 & 0.573 & 0.557  & 0.543 &  0.555 &  0.566 \\
Mg     & 0.281 & 0.242 & 0.294 & 0.279  & 0.268 &  0.279 &  0.279 \\
Al$^+$ & 0.691 & 0.643 & 0.706 & 0.691  & 0.677 &  0.688 &  0.691 \\
\hline
MAE    &       & 0.028 & 0.017 & 0.005  & 0.015 &  0.005 &  0.009 \\
\hline
\hline 
\end{tabular}
\end{table*}
The results are collected in Table \ref{tab:IP}. 
The most noteworthy features of these data are: 
(1) The pure RPA, though showing significant improvement over the
EXX-only approximation, generally overestimates the true IPs;
(2) Correspondingly, the RPA+SOX underestimates IPs (consistent with 
the unscreened nature of the pure SOX term); 
(3) Both the RPA+ and the RPA+RSOX significantly improve over the pure 
RPA results, and are even more accurate than BLYP, the ``most accurate''
standard GGA.

\section{Summary and Conclusions}
In this work, we provide benchmark results for the RPA and three simple
extensions, allowing for an unambiguous assessment of the functionals 
by comparison with exact data.
Our results confirm earlier observations of the limited applicability
of the pure RPA: 
The RPA substantially overestimates correlation energies, which then 
results in an overestimation of energy differences like ionization 
potentials.
On the other hand, our results also demonstrate that already quite 
simple extensions of the RPA can be superior to standard GGAs:
Adding either short-range corrections within the LDA (RPA+) or a 
suitably 'screened' second order exchange contribution (RPA+RSOX) 
significantly improves both absolute energies and energy differences. 
This success is consistent with the expectation that the dominant 
source of error in the RPA functional is the missing short-range 
SOX contribution.

It seems worthwhile emphasizing that the first of these extensions, 
the RPA+, essentially comes at no cost:
Compared to the computational demands of an RPA-calculation, the cost
of the LDA for the non-RPA correlation is irrelevant.
Moreover, a systematic improvement of the RPA+ by inclusion of gradient
corrections for the non-RPA correlation contributions suggests itself.
The RPA+RSOX involves an evaluation of the orbital-dependent SOX term, 
which is computationally almost as demanding as the calculation of the 
RPA energy itself, but still much less expensive than that of the fully 
screened SOX term.

In this work also several technical aspects of RPA-calculations have 
been studied systematically, of which two should be relevant beyond
the regime of atoms considered here.
The first of these aspects is the sensitivity of the RPA-expression to
the orbitals and eigenvalues used for its evaluation.
It turned out that the character of the KS spectrum inserted into
the RPA has little impact on the resulting energy.
Even if KS states obtained by LDA calculations are used the deviations 
from more accurate data remain small.
In order to cover systems with more than one occupied shell, we have
developed a partioning scheme for the frequency integration inherent
in all RPA-type functionals, which, together with a suitable
transformation of the integration variable, allows to perform the
frequency integration with a minimum number of mesh points.
Both these numerical techniques should be particularly helpful for 
applications to more complex systems.

\begin{acknowledgments}
Financial support by the Deutsche Forschungsgemeinschaft (grant EN 
265/4-2) is gratefully acknowledged. 
The calculations for this work have been performed on the computer 
cluster of the Center for Scientific Computing of 
J.W.Goethe-Universit\"at Frankfurt. 
\end{acknowledgments}

\bibliography{dft}

\end{document}